\documentstyle[12pt,prd,aps,epsf,preprint]{revtex}
\tightenlines
\draft

\begin{document}
\begin{titlepage}

\vskip 0pt plus 0.4fill

\begin{center}
    \textbf{\LARGE Angular Inflation from Supergravity}
\end{center}

\begin{center}
{\large
G.~Germ{\'a}n$^a$%
\footnote{E-mail: gabriel@fis.unam.mx},
Anupam Mazumdar$^b$ \footnote{E-mail: anupamm@ictp.trieste.it},
and 
A. P\'erez-Lorenzana$^{b,c}$%
\footnote{E-mail: aplorenz@ictp.trieste.it}\\[0.3cm]
}
\textit{
$^{(a)}$Centro de Ciencias F\'{\i}sicas,
Universidad Nacional Aut{\'o}noma de M{\'e}xico,\\
Apartado Postal 48-3, 62251 Cuernavaca, Morelos,  M{\'e}xico\\[0.3cm]
$^{(b)}$The Abdus Salam International Centre for Theoretical Physics, 
     I-34100, Trieste, Italy\\[0.3cm]
$^{(c)}$Departamento de F\'{\i}sica,
Centro de Investigaci\'on y de Estudios Avanzados del I.P.N.\\
Apdo. Post. 14-740, 07000, M\'exico, D.F., M\'exico}

\vspace{1ex}
\vskip 1ex plus 0.3fill

{\large \today}

\vskip 1ex plus 0.7fill

\textbf{Abstract}
\begin{quotation}
We study supergravity inflationary models where inflation is produced 
along the angular direction. For this we express the scalar component 
of a chiral superfield in terms of the radial and the angular components.
We then express the supergravity potential in a form particularly simple for
calculations involving polynomial expressions for the superpotential and
K\"{a}hler potential. We show for a simple Polonyi model 
the angular direction may give rise to a stage of
inflation when the radial field is fixed to its minimum. We obtain
analytical expressions for all the relevant inflationary
quantities and discuss the possibility of supersymmetry breaking 
in the radial direction while inflating by the angular component.
  \end{quotation}
  \end{center}

\vskip 0pt plus 2fill

\setcounter{footnote}{0}

\end{titlepage}

\section{Introduction}\label{intro}

Inflationary potentials in supergravity models typically involve the radial
direction of the scalar component of chiral superfields. Here we would
like to explore the possibility of having inflation along the angular 
direction. We show for superpotentials and K\"{a}hler potentials of the
polynomial type the supergravity potential can be written in
a form containing a $\cos[\phi]$-term where the field $\phi$ defines
the angular part of the scalar component of the chiral superfield of
the theory.
In particular we study a potential of the form $V=\Delta^4 (1-\cos[\phi])$.
This expression is reminiscent to pseudo Nambu-Goldstone boson (PNGB)
models of inflation dubbed as ``natural'' inflation \cite{freese}. These models
have also been used to produce curvature perturbations~\cite{lyth} by
a field other than inflaton, thus reviving some of the interesting models 
of inflation stemming from particle physics which are overrestricted by the 
number of constraints which may rule out the models from the present 
observations. One should also keep in mind that in supergravity the 
only physical scale is the Planck scale. Thus, the similarity to
natural inflation is just a formal one. 

In this paper we show that potentials of cosine type 
can be accounted for very naturally within supergravity theories, where
the field $\phi$ can play a role of inflaton in some models, or,
it could work as a curvaton in others. This leaves the possibility 
of looking for supersymmetry breaking in the radial direction in the 
first case, or, constructing models of inflation in the radial direction 
while the curvature perturbations are generated by the
angular part in the second case. In this paper we do not study all
possibilities which can occur but simply show how supergravity allows
one of the scenarios. We work out a very simple case which could
be described as a Polonyi potential~\cite{polonyi}  with an angular
component. We discuss a very nice feature of the model which
containing only one complex scalar field providing, nevertheless,
the possibility of supersymmetry breaking with a vanishing
vacuum energy in the radial direction while inflating by the angular
component. Finally, we provide analytical expressions for inflationary
quantities such as the end of inflation, scale of inflation, 
reheat temperature, number of e-folds and spectral index.

\section{The Polonyi Potential with an Angular Component} \label{natural}

Let us consider the supergravity potential for one chiral superfield with
scalar component $z$ and without D-terms~\cite{bailin}
\begin{equation}
 V = e^K
     \left[F^*(K_{zz^*})^{-1}F -
     3|W|^2 \right],
\label{pot0}
\end{equation}
where
\begin{equation}
 F \equiv \frac{\partial W}{\partial z} +
       \left(\frac{\partial K}{\partial z}\right) W ,\qquad
 K_{zz^*} \equiv \frac{\partial^2 K}{\partial z \partial z^*}~.
\end{equation}
The reduced Planck mass $M\sim 2.4 \times 10^{18}$ GeV has been set equal 
to one. The superpotential and K\"{a}hler potential denoted $W$ and $K$ 
respectively. Here we are interested in models where $W$ and $K$ are
given by polynomial expressions such as
\begin{equation}
W=\sum_{n=0}^{\infty}a_n z^n,
\label{superpot}
\end{equation}
and 
\begin{equation}
K=\sum_{n=1}^{\infty}b_n (zz^*)^n,
\label{kapot}
\end{equation}
where $a_n$ and $b_n$ are real coefficients. In general, as we have 
shown in the Appendix this structure leads to expressions that contain
$\cos$-form potentials for the angular field $\phi$ which is a real 
field defined from $z$ in a following way
\begin{equation}
z=\chi e^{i\phi}~.
\label{zeta}
\end{equation}
Here $\chi$ represents the radial field, and $\phi$ is measured in units
of reduced Planck scale. With the above definitions we claim that it is 
possible to obtain a bout of inflation. In order to illustrate this, let 
us consider the case where the superpotential is simply given by
\begin{equation}
W = a_0 + a_1 z~,
\end{equation}
with canonical kinetic energy term for which $K = z z^* = \chi^2$.
It is then straightforward to see that the supergravity potential becomes
\begin{equation}
 V = e^{\chi^2}\left(a_0^2 (\chi^2-3)+a_1^2 (1-\chi^2+\chi^4)
 +2 a_0 a_1 \chi(\chi^2-2)\cos[\phi]\right).
\label{pot4}
\end{equation}
Choosing $a_0=m^2 \beta$ and $a_1=m^2$ this is none other than the
Polonyi potential with an angular component \cite{polonyi}. For $\phi=0$ this
reduces exactly to the Polonyi case where $\beta=2-\sqrt{3}$ and
at the minimum $\chi$ takes the value $\chi_0=\sqrt{3}-1$.
In this case setting $\chi$ to its minimum the supergravity potential 
reduces to 
\begin{equation}
V = \Delta^4 (1-\cos[\phi]),
\label{potcos}
\end{equation}
where
\begin{equation}
 \Delta = (8(7-4\sqrt{3})e^{4-2\sqrt{3}})^{1/4}m \approx m~.
\label{delta}
\end{equation}
The Polonyi potential breaks supersymmetry  with
vanishing vacuum energy. It is then natural to ask whether
it is possible to break supersymmetry in the $\chi$-direction 
while inflating at the supersymmetry breaking scale along
the $\phi$-direction.
Note  that inflation comes only from the angular component,
as long as the radial direction remains fixed at its local minimum. 
On the contrary, no inflation  occurs
if the $\chi$ field  rolls down 
towards any of the local minima while keeping the angular component fixed. 
The actual dynamics of the fields including  whether inflation occurs or
not clearly depends on the  initial conditions. 
We will proceed with our study by analyzing the 
particular case where only the $\phi$ field rolls down. 
Thus, hereon we will implicitly assume that for some reason 
the radial direction has been already fixed.

\section{Angular Inflation}\label{inflation}

Here we obtain closed forms for all the relevant quantities involved
in the inflationary era for a potential Eq.~(\ref{potcos}).

1) {\bf The end of inflation.} In the model under consideration
inflation is generated while $\phi$ rolls down
from close to $\pi$ towards
the minimum at $\phi=0$, see Fig.1, and $\chi$ is fixed to its minimum
at $\chi =\sqrt{3}-1$. The end of
inflation occurs at $\phi =\phi _{e}$ when the slow roll conditions are
violated. The slow-roll conditions are upper
limits on the normalized slope and curvature of the potential, these are
given by~\cite{turner}
\begin{equation}
 \epsilon \equiv \frac{1}{6}\left(\frac{V'}{V}\right)^2 \ll 1 ,\qquad
 |\eta| \equiv \frac{1}{3}\left|\frac{V''}{V}\right| \ll 1.
\label{slowroll}
\end{equation}
The potential determines the Hubble parameter during inflation as
$H_{inf} \equiv\dot{a}/a\simeq\sqrt{V/3}$.  
Inflation ends when $\epsilon$ and/or $|\eta|$ become of ${\cal O}(1)$. 
In the model under consideration the
end of inflation is given by the saturation of the first condition in
Eq.~(\ref{slowroll}), we obtain
\begin{equation}
\phi _{e}=2~{\rm ArcTan} \left( \frac{1}{\sqrt{6}}\right).
\label{fie}
\end{equation}

2) {\bf Number of e-folds.} The number of e-folds from $\phi_{\rm H}$ to the
end of inflation at $\phi_e$ is
\begin{equation}
N_{\rm H} \equiv -\int_{\phi_{H}}^{\phi_e}
         \frac{V({\phi})}{V'({\phi})} d \phi=
         -2 \ln \left(\frac{\cos[\frac{\phi_{\rm H}}
         {2}]}{\cos[\frac{\phi_{e}}{2}]}\right),
\label{eneh}
\end{equation}
where the subscript $\rm H$ denotes the epoch at which a fluctuation of
wavenumber $k$ crosses the Hubble radius $H^{-1}$ during inflation,
i.e. when $aH=k$. (We normalize $a=1$ at the present epoch, when the
Hubble expansion rate is $H_0\equiv 100 h$~km\,s$^{-1}$Mpc$^{-1}$, with
$h\sim0.5-0.8$).
{}From Eqs.~(\ref{fie}) and (\ref{eneh}) we get
\begin{equation}
\phi_{\rm H}= 2~{\rm ArcCos}\left[\sqrt{\frac{6}{7}}~e^{-\frac{N_{\rm H}}{2}}\right].
\label{fih}
\end{equation}

3) {\bf Scalar density perturbations.} The adiabatic scalar density
perturbations are  generated through quantum fluctuations of the inflaton
field. The amplitude of the perturbations is measured by ~\cite{review}
\begin{equation}
\delta_{\rm H} (k) = \frac{1}{\sqrt{75}\pi} 
\frac{V_{\rm H}^{3/2}}{V_{\rm H}^{'}}.
\label{deltah}
\end{equation}
Solving Eq. (\ref{deltah}) together with Eq. (\ref{fih}) we find
\begin{equation}
A _{\rm H}=\sqrt{2} \Delta^2\left[\sqrt{\frac{7}{6}}~e^{\frac{N_{\rm H}}{2}}-
\sqrt{\frac{6}{7}}~e^{-\frac{N_{\rm H}}{2}}\right],
\label{Ah}
\end{equation}
where $A_{\rm H}\equiv \sqrt{75}\pi \delta _{\rm H}.$
The COBE observations \cite{cobe} of anisotropy in
the cosmic microwave background on large angular-scales provide
\begin{equation}
 \delta_{COBE} \simeq 1.9 \times 10^{-5} ,
\label{cobe}
\end{equation}
on the scale of the observable universe
($k_{COBE}^{-1}\sim\,H_0^{-1}\sim3000h^{-1}$~Mpc). In addition the
COBE data fixes the spectral index,
$n_{\rm H}(k)\equiv 1+\partial\ln\delta_{\rm H}^2(k)/\partial\ln k$, 
which is usually written as
\begin{equation}
n_{\rm H}(k)=1-6\epsilon_{\rm H}+2\eta_{\rm H}
=-\frac{4}{3}~\frac{\cos[\phi_{\rm H}]}{1-\cos[\phi_{\rm H}]}\approx 
\frac{2}{3}~.
\label{si}
\end{equation}
Notice that $n_{\rm H}$ is independent of the scale of inflation.

4) {\bf Reheat temperature.}
The reheat temperature at the beginning of the radiation-dominated era
is given by
\begin{equation}
T_{rh} \approx \left( \frac{90}{\pi^2g_{*}}\right)^{\frac{1}{4}}\sqrt{\Gamma},
\label{trh}
\end{equation}
where $\Gamma$ is the decay rate of the $\phi$ field.
When $\phi$ decays with gravitational
strength interactions only this is given by $\Gamma\approx \lambda^2\Delta^6$,
where $\lambda$ is the coupling constant.
In the expression for the reheat temperature above 
$g_*$ is the number of relativistic
degrees of freedom which for the minimal supersymmetric standard model 
gives $915/4$.

5) {\bf Scale of inflation.}
We approximate Eq.~(\ref{Ah}) by throwing away the second term
in the bracket which is exponentially suppressed for large values of $N_H$. 
Thus we have
\begin{equation}
A _{\rm H}=\sqrt{\frac{7}{3}}~\Delta^2 e^{\frac{N_{\rm H}}{2}}.
\label{Ahapprox}
\end{equation}
This equation determines $\Delta$ once the number of e-folds is specified.
The number of $e$-folds $N_{\rm H}$ is given by
\cite {number}
\begin{equation}
N_{\rm H}\approx 67+\frac{1}{3}\ln H+\frac{1}{3}\ln T_{rh}.
\label{nefolds}
\end{equation}
Solving the above equation consistently with Eq.~(\ref{Ahapprox}),
we obtain
\begin{equation}
\Delta=\left(\sqrt{\frac{3}{7}}
A _{\rm H} e^{-\frac{N_{\lambda}}{2}}\right)^{\frac{6}{17}},
\label{Delta1}
\end{equation}
where
\begin{equation}
N _{\lambda}=67+\frac{1}{6}\ln\left(\frac{2}{3}\right)+
\frac{1}{3}\ln\left(\lambda~
\left(\frac{90}{\pi^2 g_*}\right)^{\frac{1}{4}}\right).
\label{enehlambda}
\end{equation}

We notice that the scale of inflation $\Delta$ only depends on the
coupling constant $\lambda$. Once this is given, then all relevant inflationary
parameters like the scale of inflation, reheat temperature,
spectral index and the number of e-folds can be obtained from
Eqs.  (\ref{Delta1}), (\ref{trh}), (\ref{si}), and
(\ref{nefolds}), respectively. Also note that the end of inflation is 
already fixed by the numerical value determined by Eq.~(\ref{fie}).
The upper bounds to these parameters for $\lambda=1$,
are given by $T_{rh}\approx 0.1$ GeV,  $ \Delta \approx 1.1\times10^{12}$ GeV,
$n_H\approx 0.67$ and   $ N_H\approx 42$. 
As $\lambda$ gets smaller all these quantities decrease but the spectral index
remains fixed. The possibility of having supersymmetry breaking at a scale
$10^{10}-10^{11} GeV$ induced by $\chi$ vacuum while the $\phi$ field is
inflating is not favoured by these values. It could be however that this
interesting possibility may be realized in a more elaborated model of this type.

6) {\bf Quantum fluctuations.} The value $\phi_{\rm H}$ at the beginning of
inflation should exceed the quantum fluctuations
of the inflaton $\delta\phi \approx \frac{H}{2\pi} \approx \frac{\Delta^2}
{2\pi\sqrt{3}}$. For $\phi_{\rm H}\leq \pi$ we can impose an upper bound on
\begin{equation}
\frac{\delta\phi}{\phi_{\rm H}} \approx  \frac{\Delta^2}{2 \sqrt{3} \pi^2}
\leq  10^{-8}.
\label{qf}
\end{equation}
Thus we see that $\delta\phi$ is always much less than the value of the
inflaton at the beginning of inflation which at most should start at
$\phi=\pi$.

\section{Conclusions}\label{con}

We have argued that a supergravity potential
involving polynomial expressions for
the superpotential and the K\"{a}hler potential depending
on a single complex scalar field $z$ can in general contain $\cos[\phi]$-terms
coming from the angular direction. 

We work out in detail the simplest possible model which gives rise to this
type of potential: a Polonyi potential with an angular component. 
This model is interesting because it raises
the possibility of having supersymmetry breaking with vanishing vacuum energy
in the radial direction while
inflating in the angular one. The results obtained in this simple model,
however, does not favour the model because the spectral index comes out to be 
particularly low. We study the
angular direction of the potential and
obtain closed expressions for all the quantities relevant during
inflation such as the end of inflation, scale of inflation, reheat temperature,
spectral index and number of e-folds. We also argue that the potential
in the $\phi$-direction could be used instead as a model for the curvaton
leaving the radial part as an inflationary sector, though we do not
discuss more on this here. 


\section*{Appendix}

Here we obtain an expression for the supergravity potential particularly
appropriated for calculations involving polynomial expressions for
the superpotential and K\"{a}hler potential. Our main goal is to show the 
appearance of $\cos$-type terms on the angular field potential.

By using the  superpotential and
K\"{a}hler potential as  given by Eqs.~(\ref{superpot}) and (\ref{kapot}), 
it is straightforward to show that the supergravity potential can be written
in the form
\begin{equation}
 V = e^K
     \sum_{n=0}^{\infty} \sum_{m=0}^{\infty}\left[\frac{(n+K_1)(m+K_1)}{K_2} -
     3 \right]a_n a_m z^n z^{*m},
\label{pot1}
\end{equation}
where $K_i$ denote the sums
\begin{equation}
K_1=\sum_{n=1}^{\infty}nb_n (zz^*)^n~,
\label{kapot1}
\end{equation}
\begin{equation}
K_2=\sum_{n=1}^{\infty}n^2b_n (zz^*)^n~.
\label{kapot2}
\end{equation}
Notice that for superpotentials  and  K\"{a}hler potentials of the
form Eq.~(\ref{superpot}) and Eq.~(\ref{kapot}), respectively, Eq. (\ref{pot1})
is entirely equivalent to the supergravity potential given by Eq. (\ref{pot0}).
Let us now insert the radial and angular fields by writing  
$z$ in the way expressed by Eq.~(\ref{zeta}), $z=\chi e^{i\phi}$.
The  potential is then given by
\begin{equation}
 V = e^K
     \sum_{n=0}^{\infty} \sum_{m=0}^{\infty}\left[\frac{(n+K_1)(m+K_1)}{K_2} -
     3 \right]a_na_m\chi^{n+m}\cos[(n-m)\phi],
\label{pot2}
\end{equation}
which can finally be rewritten as
\begin{equation}
 V = e^K S\left[1+\frac{2}{S}
      \sum_{m=0}^{\infty} \sum_{n>m}^{\infty}\left[\frac{(n+K_1)(m+K_1)}{K_2} -
     3 \right]a_na_m\chi^{n+m}\cos[(n-m)\phi]\right],
\label{pot3}
\end{equation}
where
\begin{equation}
S=\sum_{n=0}^{\infty}\left[\frac{(n+K_1)^2}{K_2}-3 \right]a_n^2\chi^{2n}~.
\label{sum0}
\end{equation}
Notice that indeed $\cos[\phi]$ pieces
appear in this general potential as we
have mentioned already before. However, we should mention that 
whether these terms  would give rise to inflation or not
is in fact a model dependent issue that has to be settled down 
case by case.

\section*{Acknowledgements}
G.G. was supported by the project PAPIIT IN110200,
from the National University of Mexico (UNAM).

\begin{figure}
\centerline{
\epsfxsize=300pt 
\epsfbox{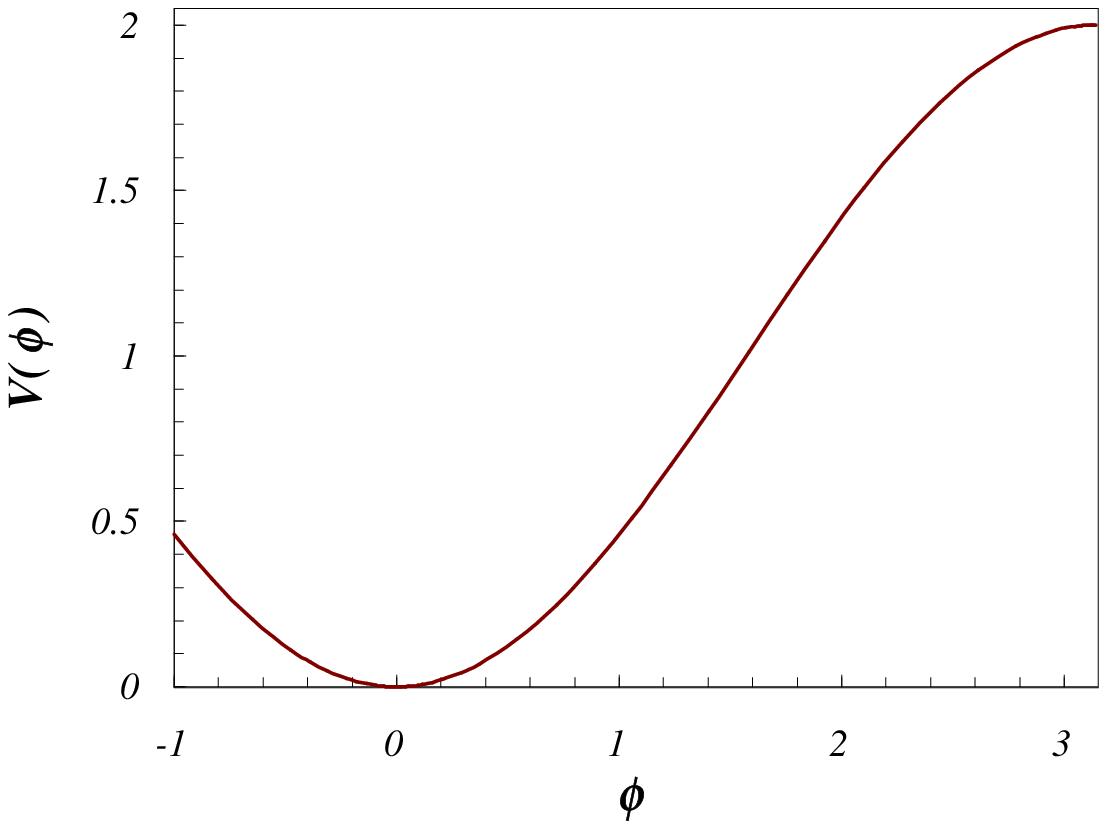}
}
\caption{The inflationary potential $V = \Delta^4 (1-\cos\phi)$
(in units of $\Delta^4$) Eq.~(\ref{potcos}),
is shown as a function of $\phi$, the angular component of the scalar field
$z$.} 
\label{fig2}
\end{figure}

\end{document}